\documentclass[usenatbib,referee]{mn2e}
\usepackage{graphicx}
\usepackage{epsfig}
\usepackage{color}

\title[Progenitors of AIC events]
{The single-degenerate model for the progenitors of accretion-induced collapse events}
\author[Wang]
{Bo Wang\thanks{E-mail:wangbo@ynao.ac.cn} \\
Yunnan Observatories, Chinese Academy of Sciences, Kunming 650216, China\\
Key Laboratory for the Structure and Evolution of Celestial Objects, Chinese Academy of Sciences, Kunming 650216, China\\
Center for Astronomical Mega-Science, Chinese Academy of Sciences, Beijing 100012, China\\
University of Chinese Academy of Sciences, Beijing 100049, China\\}

\begin{document}
\date{Accepted. Received}
\pagerange{\pageref{firstpage}--\pageref{lastpage}} \pubyear{2018}
\maketitle
 
\label{firstpage}

\begin{abstract}

It has been suggested that the accretion-induced collapse (AIC) of an oxygen-neon white dwarf (ONe WD) 
to a neutron star is a theoretically predicted outcome in stellar evolution,
likely relating to the formation of some  neutron star systems.
However, the progenitor models of AIC events are still not well studied,
and  recent studies indicated that CO WD+He star systems may also 
contribute to the formation of neutron star systems through AIC  process
when  off-centre carbon ignition happens on the surface of the CO WD.
In this work, I studied the single-degenerate (SD) model of AIC events in a systematic way, 
including  the contribution of the CO WD+He star channel and the ONe WD+MS/RG/He star channels.
Firstly, I gave the  initial parameter space of these SD channels for producing AIC events in the orbital 
period--secondary mass plane based on detailed binary evolution computations.
Then,  according to a  binary population synthesis approach, 
I gave the rates and delay times of AIC events for these SD channels based on their initial parameter space.
I found that the rates of AIC events in our galaxy are in the range of $\sim0.3-0.9\times10^{-3}$\,yr$^{-1}$,
and that their delay times are $>$30\,Myr.
I also found that the ONe WD+He star channel is the main way to produce AIC events,
and that the CO WD+He star channel cannot be ignored when studying the progenitors of AIC events.

\end{abstract}


\section{Introduction}

Carbon-oxygen white dwarfs (CO WDs) in binaries are thought to 
produce type Ia supernovae (SNe) 
when they grow in mass close to the Chandrasekhar limit (${M}_{\rm Ch}$; e.g.
Hachisu, Kato \&  Nomoto 1996; Langer et al. 2000; Podsiadlowski 2010; 
Wang \& Han 2012; Maoz, Mannucci \& Nelemans 2014).
However, oxygen-neon (ONe) WDs are  expected to collapse into neutron stars through electron-capture reactions  
once they increase their masses to ${M}_{\rm Ch}$,
in which the transformation from ONe WDs to neutron stars  is referred as the process of accretion-induced collapse 
(AIC; see, e.g. Miyaji et al. 1980; Canal et al. 1990; Nomoto \& Kondo 1991).
AIC events are a kind of electron-capture SNe, the  remnants of which are neutron star systems (e.g. Miyaji et al. 1980).
They are predicted to be very faint and fast,  and thus difficult to observe;  
a small amount ($\sim$$10^{-3}\,{M}_\odot$) of  $^{56}{\rm Ni}$ synthesized 
during collapse  indicates that AIC events are  relatively dim optical transients 
 (e.g. Woosley \& Baron 1992; Dessart et al. 2006).
Piro \& Thompson (2014) suggested that  the resulting optical transient is 
considerably fainter than that of a typical type Ia SN  (by 5 mag or more) 
and lasts for a few days to a week. 

AIC may be a viable and promising way to form some troublesome neutron star systems 
that are difficult to be explained by type II core-collapse SNe
(see Canal, Isern \& Labay 1990 for a review). For example,
low-/intermediate-mass X-ray binaries (e.g. van den Heuvel 1984; Podsiadlowski, Rappaport \& Pfahl 2002),
low-/intermediate-mass binary pulsars (e.g. Nomoto \& Kondo 1991;  
Tauris, Langer \& Kramer 2012;  Liu et al. 2018a), and
millisecond binary pulsars if they are later spun up by mass accretion
(e.g. van den Heuvel 1984; Bailyn \& Grindlay 1990; 
Bhattacharya \& Van den Heuvel 1991; Chen et al. 2011; Tauris et al. 2013; Freire \& Tauris 2014).
AIC events have been proposed as a possible source of r-process elements (e.g. 
Wheeler, Cowan \& Hillebrandt 1998; Fryer et al. 1999; Qian \& Wasserburg 2007)
and a source of gravitational wave emission (e.g.  Abdikamalov et al. 2010).
They might also  relate to the formation of  rapidly spinning  magnetars,  fast radio bursts
and ultra high-energy cosmic rays like gamma-ray bursts (e.g. Dar et al. 1992; 
Usov 1992; Piro \& Kollmeier 2016; Lyutikov \& Toonen 2017; Cao, Yu \& Zhou 2018).
Moriya (2016) recently argued that gravitational waves  
from AIC events may be accompanied by radio-bright optically faint (possibly X-ray-bright) transients,  
which can be used to verify the AIC origin of the observed gravitational waves.

Although the potential importance of AIC, there has been no reported direct detection for such events and
it is still unclear about their progenitor models.
Over the past few decades, two classic kinds of progenitor models of  AIC events have been proposed, that is,
the single-degenerate (SD) model and the double-degenerate (DD) model. 
For the SD model, an ONe WD accretes H-/He-rich material from its non-degenerate companion 
that could be a main-sequence or a slightly evolved star (the ONe WD+MS channel), 
a red-giant star (the  ONe WD+RG channel), or a He star (the  ONe WD+He star channel).
The accreted material will burn into O and Ne, and accumulate onto the surface of the ONe WD. 
AIC may occur when the ONe WD increases its mass close to ${M}_{\rm Ch}$ 
(e.g. Nomoto \& Kondo 1991;  Tauris et al. 2013; Brooks et al. 2017; Wu \& Wang 2018; Ruiter et al. 2018). 
For the DD model, it involves the merger of two WDs with a combined mass larger than ${M}_{\rm Ch}$, 
the merging of which results from the gravitational wave emission (e.g. Webbink 1984;  
Iben \& Tutukov 1984; Yoon, Podsiadlowski \&  Rosswog 2007; Ruiter et al. 2018).

Previous studies on the SD model of AIC events mainly involve the ONe WD accretors.
Yungelson \& Livio (1998) suggested that the expected AIC rate is no more than 1\% of that of SNe Ia.
Tauris et al. (2013) investigated the binary computations of ONe WD+MS/RG/He star systems 
that may experience AIC and then be recycled to produce binary pulsars.  
Although these ONe WD binary systems have been investigated by Tauris et al. (2013),
they only considered the case with the initial ONe WD mass of $1.2\,M_{\odot}$.

Additionally,  recent studies  suggested that CO WD+He star systems may also contribute 
to the formation of neutron star systems through AIC  process
when the off-centre carbon ignition occurs on the surface of the CO WD 
(e.g. Brooks et al. 2016; Wang, Podsiadlowski \& Han 2017).\footnote{The 
CO WD accretors with He donor stars are thought to be one of 
the promising ways to form the observed type Ia SNe in young populations (e.g. Wang et al. 2009; Ruiter, Belczynski \& Fryer 2009).} 
Wang, Podsiadlowski \& Han (2017) recently found that off-centre carbon burning happens on the surface of the WD 
if the mass-accretion rate is above a critical value  ($\sim$$2.05\times 10^{-6}\,{M}_\odot\,\mbox{yr}^{-1}$).
Off-centre carbon burning will convert CO WDs to ONe WDs through an inward-propagating carbon flame, 
resulting in the formation of neutron stars  through AIC process if the mass-transfer continues 
(e.g. Saio \& Nomoto 1985, 1998; Nomoto \& Iben 1985; Schwab, Quataert \& Kasen 2016; Brooks et al. 2016, 2017).  
Note that off-centre carbon burning provides an alternative way to form neutron stars,
which will increase the rates of AIC events (see also Brooks et al. 2017).

Studies of the AIC event rate using binary population synthesis (BPS) should include all of these SD channels, 
especially the CO WD+He star channel that has been previously neglected.
In this work, I will study the SD model of AIC events  in a systematic way, 
including the contribution of the CO WD+He star channel and  the ONe WD+MS/RG/He star channels.
I introduce the basic assumptions and methods for binary evolution computations, and give
the initial parameter space  of AIC events for the CO WD+He star channel in Section 2
and the ONe WD+MS/RG/He star channels in Section 3.
In Section 4, I show  the BPS methods and the corresponding  results.
Finally, a discussion is given in Section 5 and a summary in Section 6.

\section{The CO WD+He star channel}
\subsection{Numerical methods}
In the CO WD+He star channel, a CO WD accretes He-rich material from a He star once it fills its Roche lobe.
The He star transfers some of its material onto the surface of the CO WD, resulting in the mass growth of the WD.
Previous studies usually assumed that a He-accreting CO WD can explode as a type Ia SN  once it grows in mass to 
${M}_{\rm Ch}$ (set to be 1.38\,${M}_\odot$; e.g. Wang et al. 2009; Ruiter, Belczynski \& Fryer 2009). 
However,   recent investigations found that for relatively high mass-accretion rates carbon can be ignited off-centre 
when a  CO WD accretes He-rich material from its companion,  probably leading to collapse into a 
neutron star but not a type Ia SN (e.g. Brooks et al. 2016; Wang, Podsiadlowski \& Han 2017). 
If the mass-accretion rate $\dot{M}_{\rm acc}$ is above a critical value 
 ($\dot{M}_{\rm cr}\sim2.05\times 10^{-6}\,{M}_\odot\,\mbox{yr}^{-1}$),  
Wang, Podsiadlowski \& Han (2017) found that
off-centre carbon burning occurs on the surface of the CO WD when it grows in mass 
close to ${M}_{\rm Ch}$. For the conditions of relatively low mass-accretion rates ($\dot{M}_{\rm acc}<\dot{M}_{\rm cr}$), 
carbon can be ignited in the centre when a  CO WD grows in mass close to ${M}_{\rm Ch}$, 
leading to a type Ia SN explosion  (see Wang, Podsiadlowski \& Han 2017).

On the basis of the optically thick wind assumption (see Hachisu, Kato \&  Nomoto 1996),\footnote{Note that 
the optically thick wind model is still under hot debate. 
For example, the metallicity  threshold predicted by this model is conflict with observations
(e.g. Prieto, Stanek \& Beacom 2008; Badenes et al. 2009; Galbany et al. 2016), and
 the wind velocity produced by this model is too large to match observations (e.g. Badenes et al. 2007; Patat et al. 2007).
In addition, some emission lines (e.g. He II 4686 and [O I] 6300) predicted by this model  has not been reported
(e.g. Woods \& Gilfanov 2013; Graur et al. 2014; Johansson et al. 2016). Note also that two objects 
(RX J0513.9$-$6951 and V Sge) are in the accretion wind evolution phase, 
which strongly supports the existence of accretion wind (see Hachisu \& Kato 2003a,b).}
Wang et al. (2009)  performed a systematic study of
the CO WD+He star channel for the progenitors of type Ia SNe.
They carried out  binary evolution computations with the Eggleton stellar evolution code (Eggleton 1973; 
Han, Podsiadlowski \& Eggleton 1994; Pols et al.\ 1998)
to estimate the initial parameter space of WD binaries  that can lead to type Ia SNe
in the orbital period--secondary mass ($\log P^{\rm i}-M^{\rm i}_2$) plane.  
However,  Wang et al. (2009)  ignored the possibility of non-explosive off-centre carbon ignition.
The effect of including this possibility will increase the initial parameter spaces for producing AIC events
in the $\log P^{\rm i}-M^{\rm i}_2$ plane.
Here, I re-used the mass-transfer histories from Wang et al. (2009), and then assume that
off-centre carbon burning happens if $\dot{M}_{\rm acc}$ is
higher than $\dot{M}_{\rm cr}$ when the mass of the WD approaches ${M}_{\rm Ch}$,
resulting in the occurance of AIC events.
It is worth noting that the maximum stable mass of a WD may exceed ${M}_{\rm Ch}$ 
when rotation is considered (e.g. Yoon \& Langer 2004).

\subsection{An example of binary evolution computations}

\begin{figure*}
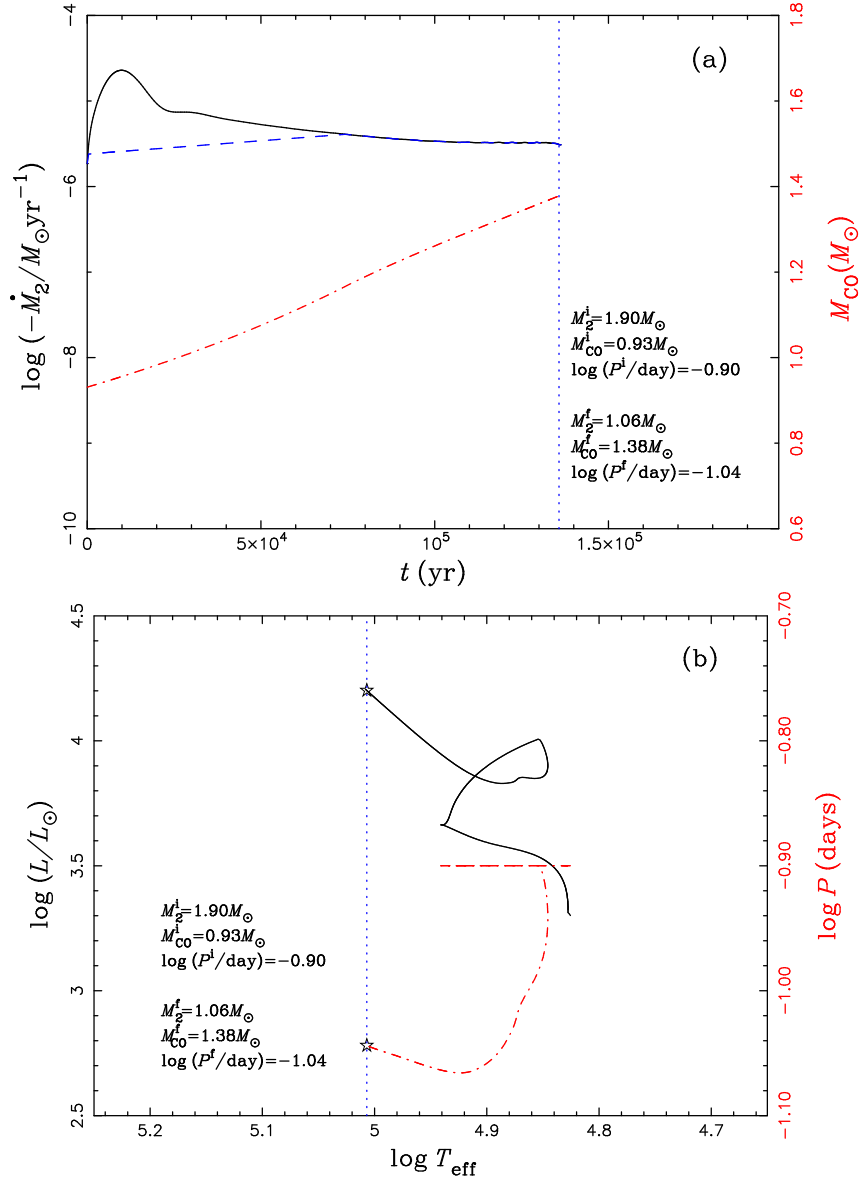

\includegraphics[width=7.8cm,angle=270]{f1a.eps}
\includegraphics[width=7.8cm,angle=270]{f1b.eps}
\caption{A representative example of binary evolution computations for the CO WD+He  star channel  to produce an AIC event.
Panel (a): 
the solid, dashed and dash-dotted curves show the mass-transfer rate, 
the mass-retention rate and the mass of the CO WD varying with time, respectively. 
Panel (b): 
the evolutionary track of the He star is shown as a solid curve and the evolution of orbital period is shown as a dash-dotted curve. 
Dotted vertical lines in both panels and stars in panel (b) indicate the position where the CO WD is expected to collapse into a neutron star. 
The initial binary parameters and the parameters at the moment of AIC are given in these two panels.}
\end{figure*}

In Fig. 1, I  present a representative example of binary evolution computations for the CO WD+He  star channel to produce an AIC event.
Panel (a)
shows the mass-transfer rate, the mass-retention rate and the mass of the CO WD varying with time after the He donor star fills its Roche lobe,
and panel (b) presents the evolutionary track of the He
donor star in the Hertzsprung-Russell diagram, in which the evolution of the orbital period is also presented.
The CO WD+He  star  binary presented in this case is ($M_2^{\rm i}$, $M_{\rm CO}^{\rm i}$, $\log
(P^{\rm i}/{\rm d})$) $=$ (1.90, 0.93, $-$0.90), in which $M_2^{\rm i}$,
$M_{\rm CO}^{\rm i}$ and $P^{\rm i}$ are the initial masses of the He donor
star and of the CO WD in ${M}_\odot$, and the initial orbital
period in days, respectively. 

The mass-transfer rate $\dot{M}_{\rm 2}$ is larger than the maximum  accretion rate for stable He-shell burning
 soon after the onset of Roche-lobe overflow, leading to a wind phase,
during which a part of the transferred mass is blown off in the form of the optically
thick wind, and the left material is accumulated onto the CO WD. After about
$7.0\times10^{4}$\,yr, $\dot{M}_{\rm 2}$ drops below the maximum  accretion rate
 but still higher than the minimum  accretion rate for stable He-shell burning. At this stage, the optically
thick wind stops and the He-shell burning becomes stable. After about $6.5\times10^{4}$\,yr, the CO WD
grows in mass to $1.38\,M_{\odot}$, and is expected to collapse into a neutron star due to off-centre carbon burning. 
At this moment, the mass of the
donor star is $M^{\rm f}_2=1.06\,M_{\odot}$ and the orbital
period $\log (P^{\rm f}/{\rm day})=-1.04$.

\subsection{Parameter space for AIC events}

\begin{figure}
\begin{center}
\epsfig{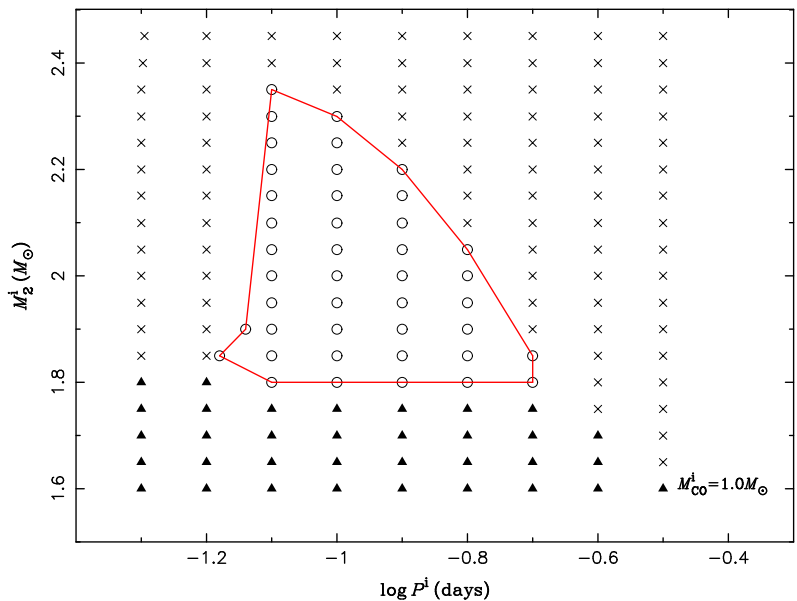} 
\caption{Initial parameter space of AIC events 
in the $\log P^{\rm i}-M^{\rm i}_2$ plane  for ${M}^{\rm i}_{\rm CO}=1.0\,M_{\odot}$.
Open circles are those under off-centre carbon ignition, resulting in the formation of neutron star systems via the AIC process finally.
The filled triangles denote systems that lead to type Ia SN explosions. 
Crosses denote systems that experience strong He-shell flashes, preventing WDs from growing in mass to
${M}_{\rm Ch}$.}
\end{center}
\end{figure}

Fig. 2 presents the final results
of the binary evolution computations in the $\log
P^{\rm i}-M^{\rm i}_2$ plane for an initial CO WD mass ${M}^{\rm i}_{\rm CO}=1.0\,M_{\odot}$, 
where open circles are those under off-centre carbon ignition that results in the formation of neutron star systems finally. 
The filled triangles denote CO WD+He star  systems that lead to type Ia SN  explosions but not AIC events.
The crosses denote systems that experience strong He-shell flashes, preventing WDs from growing in masses to
${M}_{\rm Ch}$.

\begin{figure}
\begin{center}
\epsfig{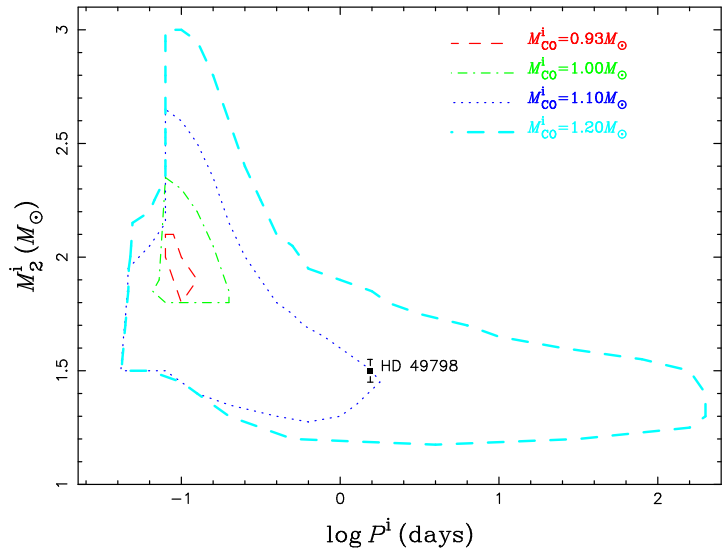} \caption{Initial parameter space of AIC events for the CO WD+He star channel
in the $\log P^{\rm i}-M^{\rm i}_2$ plane with different ${M}^{\rm i}_{\rm CO}$.  
The location of HD 49798 with its WD companion is indicated in this figure.}
\end{center}
\end{figure}

Fig. 3 shows the initial contours of AIC events for the CO WD+He star channel
in the $\log P^{\rm i}-M^{\rm i}_2$ plane with various ${M}^{\rm i}_{\rm CO}$ (i.e.
$M_{\rm CO}^{\rm i}$ = 0.93, 1.0, 1.1 and $1.2\,M_{\odot}$).
If  the initial parameters of a CO WD+He star system are 
located in this parameter space, an AIC event  is predicted to be formed.
The minimum $M_{\rm CO}^{\rm i}$ for producing AIC events in this channel is 
$0.93\,M_{\odot}$ that can grow in mass to ${M}_{\rm Ch}$.
The binary parameters of HD 49798 with its WD companion are located in 
the parameter space of WD+He star systems for producing AIC events (see Fig. 3).
Thus, I speculate that
HD 49798 with its WD companion may 
be a progenitor candidate of AIC events (for more discussions see Sect. 5).

In Fig. 3,
the upper boundaries of the parameter space are mainly determined  by a high $\dot{M}_{\rm 2}$ due to a large mass-ratio, 
which makes the CO WD+He star systems lose too much mass through the optically thick wind, preventing CO WDs from
increasing their masses to ${M}_{\rm Ch}$.
The left-hand boundaries  are mainly constrained 
 by the minimum value of $\log P^{\rm i}$, for which a zero-age He MS star would fill its Roche lobe.
CO WD+He star systems below the lower boundaries could produce type Ia SNe (see Wang et al. 2009). 
Systems beyond the right-hand boundaries experience a relatively 
high $\dot{M}_{\rm 2}$ because of the rapid expansion of He stars when they evolve to the 
subgiant phase, some of which may contribute to 
the formation of double CO WDs and produce type Ia SNe finally (e.g. Ruiter et al. 2013; Liu et al. 2016; Liu, Wang \& Han 2018).

\section{The ONe WD+MS/RG/He star channels}
\subsection{Numerical methods}
ONe WDs are expected to collapse into neutron stars
through AIC process once they increase their masses to ${M}_{\rm Ch}$. 
In the SD model, an ONe WD could accrete H-/He-rich material from a non-degenerate companion 
that could be a MS or a slightly evolved star (the MS donor channel), 
a red-giant star (the  RG donor  channel), or a He star (the  He star donor  channel).
The accreted material will burn into O and Ne, and accumulate onto the surface of the ONe WD. 
The electron capture reactions will be triggered in the centre of the ONe WD when it
grows in mass to ${M}_{\rm Ch}$, leading to  the occurance of AIC events (e.g. Wu \& Wang 2018).

According to the optically thick wind assumption, our previous studies have already obtained a dense model
grid leading to  WDs  with ${M}_{\rm Ch}$ for various initial WD masses ${M}^{\rm i}_{\rm ONe}$ at metallicity $Z=0.02$
except for ${M}^{\rm i}_{\rm ONe}=1.3\,M_{\odot}$
 (for the MS donor channel see Wang et al. 2014, see also Han \& Podsiadlowski 2004, 2006; 
 for the RG donor channel see Liu et al. 2018b;  
 for the He star donor channel see Wang et al. 2009).
Adopting our previous assumptions, I obtained the initial parameter space leading to 
AIC events with $M_{\rm ONe}^{\rm i}=1.3\,M_{\odot}$ for various non-degenerate companions.

\subsection{Parameter space for AIC events}
\begin{figure}
\begin{center}
\epsfig{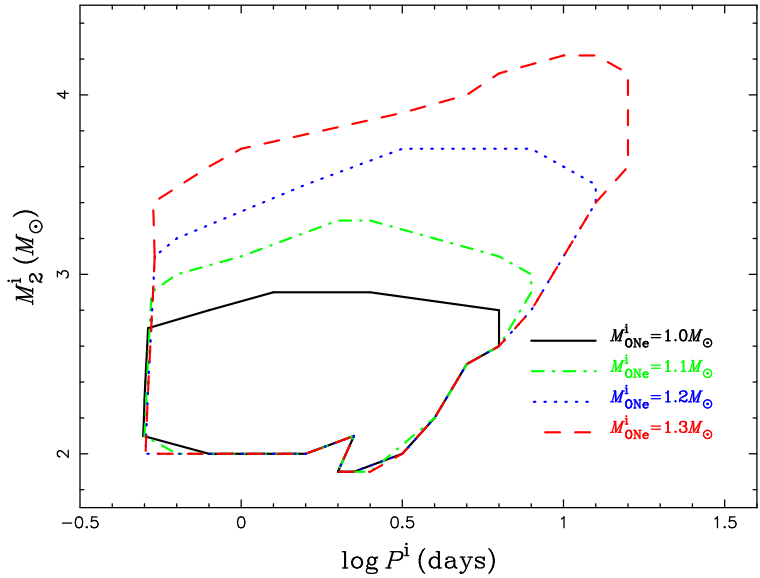} \caption{Initial parameter space of AIC events for the ONe WD+MS channel
in the $\log P^{\rm i}-M^{\rm i}_2$ plane with different ${M}^{\rm i}_{\rm ONe}$. }
\end{center}
\end{figure}

\begin{figure}
\begin{center}
\epsfig{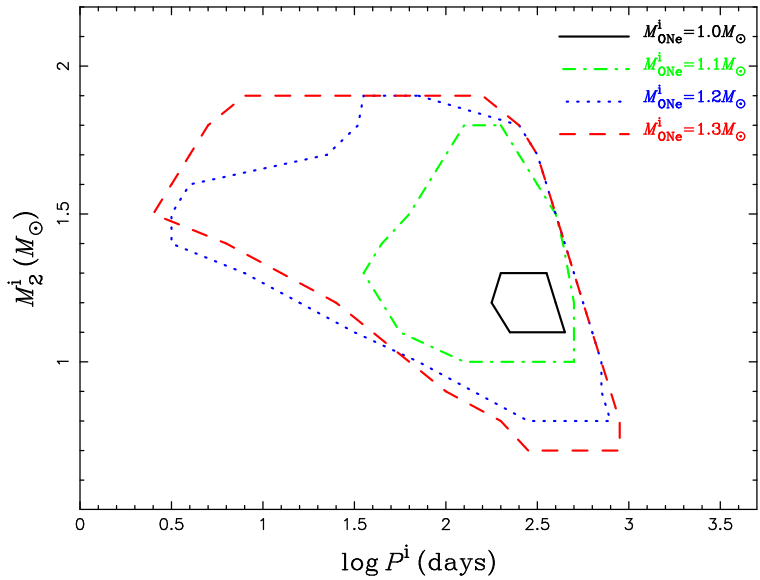} \caption{Initial parameter space of AIC events for the ONe WD+RG channel
in the $\log P^{\rm i}-M^{\rm i}_2$ plane with different ${M}^{\rm i}_{\rm ONe}$. }
\end{center}
\end{figure}

\begin{figure}
\begin{center}
\epsfig{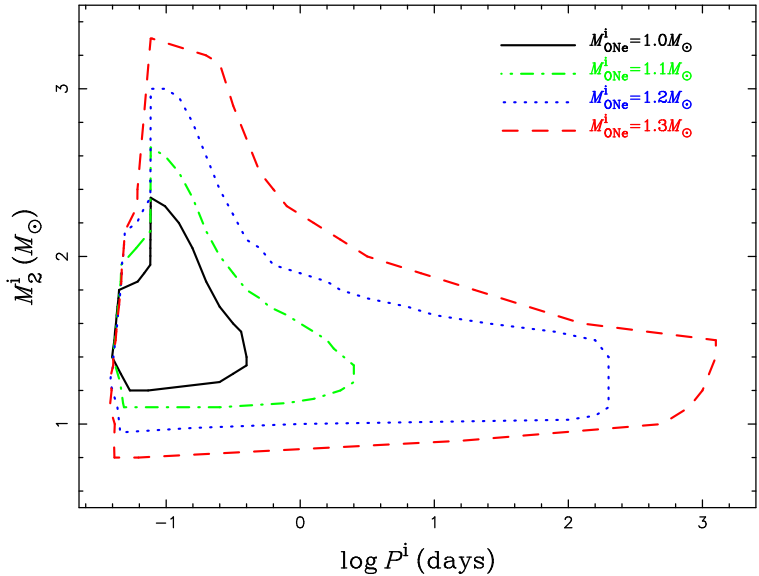} \caption{Initial parameter space of AIC events for the ONe WD+He star channel
in the $\log P^{\rm i}-M^{\rm i}_2$ plane with different ${M}^{\rm i}_{\rm ONe}$. }
\end{center}
\end{figure}

Figs 4-6 presents the initial contours of AIC events for the ONe WD+MS/RG/He star channels
in the $\log P^{\rm i}-M^{\rm i}_2$ plane with various ${M}^{\rm i}_{\rm ONe}$, respectively.
Similar to Fig. 3, the upper boundaries of these  initial contours and WD systems 
beyond the right-hand boundaries are mainly determined  by a high $\dot{M}_{\rm 2}$,
and the left-hand boundaries  are mainly constrained 
by the minimum value of $\log P^{\rm i}$, for which a zero-age MS star or He star  would fill its Roche lobe.
The lower boundaries are constrained by the condition that $\dot{M}_{\rm 2}$ 
should be high enough to ensure the mass growth of the ONe WD during H-/He-shell flashes.
Note that  the initial contours for producing AIC events may be changed if a different metallicity is used;
the initial contours would be enlarged to have larger mass donors and longer orbital periods 
in higher metallicity  environments
(e.g. Meng, Chen \& Han 2009; Wang \& Han 2010a).

\section{Binary population synthesis}
\subsection{BPS methods}

In order to obtain  AIC rates and delay time distributions (DTDs) for these SD channels, 
a series of Monte Carlo BPS simulations are performed.
For each BPS simulation, I employed the
Hurley rapid binary evolution code (Hurley, Tout \& Pols 2002) to simulate
the evolution of $10^{\rm 7}$ primordial binaries with metallicity $Z=0.02$. 
These binaries are followed from
the star formation to the formation of WD binaries (CO WD+He star systems or  ONe WD+MS/RG/He star systems;
for binary evolutionary paths of these systems see Wang 2018).

If the initial parameters of a WD binary at the onset of the
Roche-lobe overflow are
located in the  production regions of AIC events in the $\log P^{\rm i}-M^{\rm i}_2$ 
plane for its specific initial WD mass,  then an AIC event
is assumed to happen.  I adopt linear interpolation if
$M_{\rm WD}^{\rm i}$ is not among the masses listed in Figs 3--6.
The standard energy equations are adopted to calculate the output 
during the common-envelope (CE) evolution (e.g. Webbink 1984). 
Here, I set the CE ejection efficiency ($\alpha_{\rm CE}$) to be
1.0 and 3.0 to examine its influence on the final results.

Eight sets of Monte Carlo BPS simulations were conducted to study the rates of AIC events (see Table 1).
A delta function star-formation is simply modeled,  i.e. a single instantaneous starburst
($10^{10}\,M_{\odot}$ in stars is supposed). Alternatively, 
a constant star-formation rate  of  $5\,{M}_{\odot}\rm yr^{-1}$ is assumed over the past 14 Gyr.
The following basic assumptions for
these Monte Carlo simulations are adopted:
(1) All stars are  supposed to be members of binaries.
The primordial binary samples are generated through a Monte Carlo way, and
an initially circular orbit was assumed for all primordial binaries. 
(2) The distribution
of initial orbital separations is assumed to be constant in $\log a$ for
wide binaries, where $a$ is the orbital separation (e.g. Han, Podsiadlowski \& Eggleton 1995).
(3) The initial mass function  of the primordial primary star is taken
from Miller \& Scalo (1979). 
(4) A constant mass-ratio distribution is taken, that is, $n(q)=1$.

\subsection{BPS results}

\begin{table}
 \begin{center}
 \caption{Galactic rates of AIC events for different Monte Carlo BPS simulation sets. 
 Notes: 
 $\alpha_{\rm CE}$ = CE ejection efficiency; 
$\nu_{\rm AIC}$ = Galactic rates of AIC events;
${\rm DTDs}$ = Delay time distributions of AIC events.}
   \begin{tabular}{cccccccc}
\hline \hline
Set & $Channel $ & $\alpha_{\rm CE}$  & $\nu_{\rm AIC}$  & ${\rm DTDs}$\\
&&&($10^{-3}$\,yr$^{-1}$)&(Myr)\\
\hline
$1$ & ${\rm CO\,WD+He\,star}$     & $1$    & $0.083$      & $50-110$ \\
$2$ & ${\rm CO\,WD+He\,star}$     & $3$   & $0.129$      & $50-110$ \\
$3$ & ${\rm ONe\,WD+MS}$          & $1$    & $0.138$      & $110-1400$ \\
$4$ & ${\rm ONe\,WD+MS}$          & $3$   & $0.079$      & $70-1400$ \\
$5$ & ${\rm ONe\,WD+RG}$          & $1$    & $0.012$      & $1400-6300$ \\
$6$ & ${\rm ONe\,WD+RG}$          & $3$     & $0.033$     & $1400-6300$\\
$7$ & ${\rm ONe\,WD+He\,star}$   & $1$     & $0.105$     & $40-140$ \\
$8$ & ${\rm ONe\,WD+He\,star}$   & $3$     & $0.676$     & $30-180$ \\
\hline
\end{tabular}
\end{center}
\end{table}

\begin{figure}
\begin{center}
\epsfig{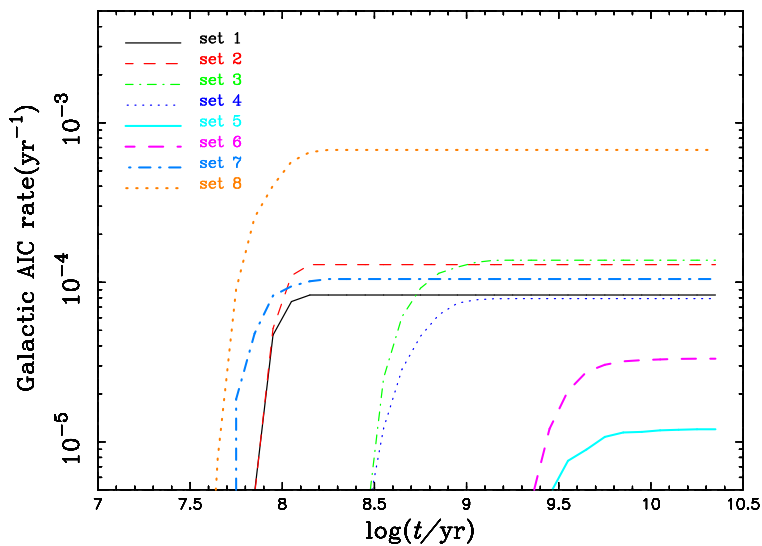}
\caption{Rates of  AIC events  in the Galaxy for a constant star-formation rate of $5\,M_{\odot}$\,yr$^{-1}$. }
\end{center}
\end{figure}

I performed eight sets of simulations  with metallicity $Z=0.02$
to systematically investigate the Galactic rates of AIC events for these SD channels, including
the CO WD+He star channel and the ONe WD+MS/RG/He star channels (see Table 1).
I vary the values of $\alpha_{\rm CE}$ to examine its influence on the final results.

In Fig. 7, I show the evolution of AIC rates changing with time by adopting metallicity $Z=0.02$ and 
star-formation rate of $5\,{M}_{\odot}\rm yr^{-1}$. 
This study presents  the rates of AIC events in the Galaxy are
$\nu_{\rm AIC}\sim0.3-0.9\times10^{-3}$\,yr$^{-1}$.
I found that the ONe WD+He star channel is the main way to contribute AIC events, and that
the CO WD+He star channel cannot be ignored when studying AIC events.
According to the eight sets of simulations for these SD channels, 
the estimated rates of AIC events are strongly sensitive to the choice of $\alpha_{\rm CE}$.
For example, 
AIC rates for the ONe\,WD+He\,star channel become lower with the decrease of $\alpha_{\rm CE}$ (see sets 7-8). 
This is because more binaries after the formation of the CE will merge for a lower value of $\alpha_{\rm CE}$.

The rate of AIC events is still uncertain.  
The ejecta from AIC events has been claimed as an alternative source of r-process nucleosynthesis. 
By calculating the  r-process nucleosynthetic yields  of neutron rich ejecta from AIC events, 
Fryer et al. (1999) inferred that the rates of  AIC events are roughly in the range of  $10^{-7}-10^{-5}$\,yr$^{-1}$. 
However, whether AIC process produces r-process elements is still under debate
(e.g. Fryer et al. 1999; Qian \& Wasserburg 2007).
In the present work, I give the Galactic rates of AIC events in the range of 
$\sim0.3-0.9\times10^{-3}$\,yr$^{-1}$, which at least provides an upper limit for the AIC rate from these SD channels.

\begin{figure}
\begin{center}
\epsfig{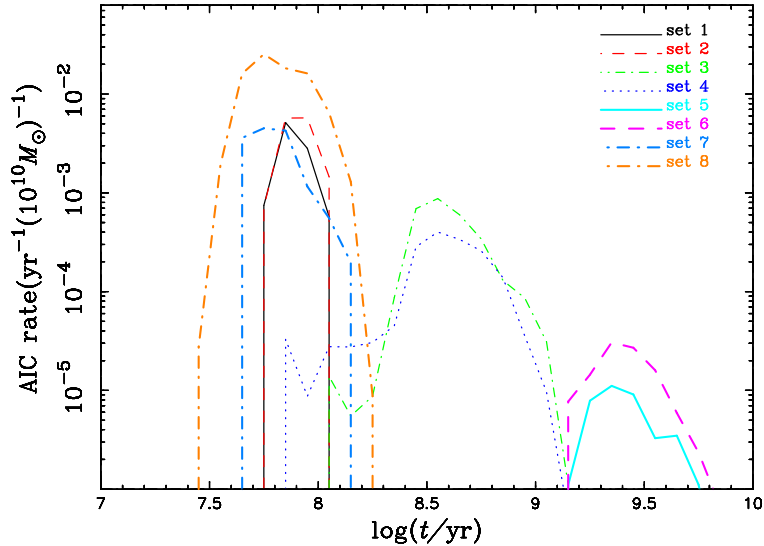}
\caption{Similar to Fig. 7,  but for a single starburst with a total mass of $10^{10}\,M_{\odot}$.}
\end{center}
\end{figure}

The delay time distributions (DTDs) of AIC events are defined as the time interval between the star formation and AIC. 
Fig. 8 presents the DTDs of AIC events based on a single starburst with a total
mass of $10^{10}\,M_{\odot}$ in stars. The delay times of AIC events are $>$30Myr after the
starburst, in which the CO/ONe WD+He star channels contribute to AIC events with short delay times,
the ONe WD+MS channel contributes to  AIC events with intermediate delay times, and
the ONe WD+RG channel contributes to  AIC events with long delay times.
The minimum delay time in Fig. 8 is determined by the MS lifetime of  the maximum  mass  of a star  that can form ONe WDs,
which can be affected by different metallicities (see Doherty et al. 2017).

\section{Discussion}

AIC  has been suggested as a natural theoretical final fate  for  ONe WDs
once their masses approach  ${M}_{\rm Ch}$. However, the final outcome of accreting ONe WDs is sill under debate.
Jones et al. (2016) found that the deflagration of oxygen could happen in the degenerate ONe cores, 
leading to  the formation of a subpopulation of type Ia SNe  (see also Miyaji et al. 1980; Isern, Canal \& Labay 1991).  
However,  Wu \& Wang (2018) recently found that the final outcome of accreting ONe WDs is 
electron-capture induced collapse rather than thermonuclear explosion 
though different initial ONe WD masses and mass-accretion rates could affect the evolution of central density and temperature; 
the central temperature of the accreting ONe WDs cannot reach the explosive oxygen or neon ignition temperature due to neutrino cooling,
resulting in the formation of neutron stars finally
(see also Schwab, Quataert \& Bildsten 2015; Brooks et al. 2017).

Tauris et al. (2013) studied the binary evolution of ONe WD$+$MS/RG/He star systems 
that may experience AIC and then be recycled to form binary pulsars. However,
they only provided a parameter space of these ONe WD binary systems for producing AIC events 
with $M^{\rm i}_{\rm ONe}=1.2\,\rm M_{\odot}$.
Compared with the results of Tauris et al. (2013),  the initial parameter space in the present work 
has more massive donor stars and longer orbital periods. 
In the work of Tauris et al. (2013), they assumed that CE will be formed if $\dot{M}_{\rm 2}$ 
is larger than a critical value (i.e. three times of Eddington mass-accretion rate).  
As Tauris et al. (2013) pointed out,  the critical value they adopted may be the largest uncertainties in their calculations. 

In the SD model, the  progenitors of AIC events could be CO WD+He star systems or ONe WD binary systems.
CO WD+He star systems could produce neutron star+He star systems if AIC happens, 
in which neutron stars may be recycled when the He stars refill their Roche lobe, 
resulting in the formation of  intermediate-mass binary pulsars finally.
ONe WD binary systems will evolve to some different neutron star systems if AIC occurs, 
depending on star types of their mass donors, as follows:
(1) The ONe WD$+$MS systems may experience AIC and 
evolve to form fully recycled millisecond binary pulsars with He WD companions (e.g. Tauris et al. 2013; Ablimit \& Li 2015).
(2) The ONe WD$+$RG systems are more likely to form intermediate-mass binary pulsars,  
including mildly recycled pulsars and CO/ONe WD companions with long orbital periods (e.g. Tauris et al. 2013). 
(3) The ONe WD$+$He star systems may experience AIC and eventually form 
intermediate-mass binary pulsars with short orbital periods (e.g. Liu et al.  2018a).

A number of WD binaries are known to be possible candidates for the SD progenitors of AIC events.
The most relevant known binary system to the present work is perhaps HD 49798 with its WD companion.
HD 49798 is a hydrogen stripped  subdwarf O6 star (1.50$\pm$0.05$\,M_{\odot}$) that has
an X-ray pulsating companion RX J0648.0$-$4418 (1.28$\pm$0.05$\,M_{\odot}$) with an orbital period of 1.548\,d,
but the nature of the  compact companion is still under debate (e.g. Thackeray 1970; 
Bisscheroux et al. 1997;  Mereghetti et al. 2009, 2016; Liu et al. 2015, 2018a;  Popov et al. 2018).
By analyzing the angular momentum and magnetic field,
Mereghetti et al. (2016) argued that the X-ray pulsating companion  of HD 49798 is likely 
a neutron star  (see also Brooks, Kupfer \& Bildsten 2017).
However, the large emitting radius ($\sim$40\,km) obtained from the black body spectral is still puzzling.
Popov et al. (2018)  recently suggested that  the contraction of a young WD can well explain
the continuous stable spin-up of the  compact companion.  Liu et al. (2018a)  also
suggested that the  companion of HD 49798 may be a WD but not a neutron star 
based on binary evolution computations.
Assuming the companion of HD 49798 is a CO WD, 
Wang \& Han (2010b) found that the massive
WD can grow in mass to ${M}_{\rm Ch}$ after about $6\times10^{4}$\,yr based on detailed
binary evolution computations.    In this binary, off-centre carbon burning will happen when the WD  increases its mass  close to ${M}_{\rm Ch}$
due to the high $\dot{M}_{\rm 2}$ ($>$$2.05\times
10^{-6}\,{M}_\odot\,\mbox{yr}^{-1}$; see Wang \& Han 2010b).  Thus, I speculate that the massive WD in this binary may
eventually form a neutron star through AIC and but not a type Ia SN. 

In addition, U Sco,  TCrB and RS Oph  contain massive WDs
that are already close to ${M}_{\rm Ch}$, which are possible progenitor candidates of AIC events  (e.g. 
Hachisu, Kato \& Nomoto 1999; Parthasarathy et al. 2007). 
U Sco is a recurrent nova,  including a $1.55\pm0.24\,M_{\odot}$ 
WD  and a $0.88\pm0.17\,M_{\odot}$ MS donor
with an orbital period of $\sim1.23$\,d (e.g. Thoroughgood et al. 2001).
Mason (2011)  suggested  that the WD in U Sco may be an ONe WD based on spectroscopic observations, and  thus its
final fate may collapse into a neutron star but not a type Ia SN. 
Note that Mason (2013) rectified her conclusion in the erratum and  concluded that there is no
evidence on neon overabundance in the ejecta of U Sco (see also Miko\l{}ajewska \& Shara 2017).
T CrB is a symbiotic system, including 
a $\sim$$1.2\,M_\odot$  WD and a $\sim$$0.7\,M_\odot$ RG star with an orbital period of $\sim$227\,d 
 (e.g. Belczy$\acute{\rm n}$ski \& Miko\l{}ajewska 1998).
 RS Oph is a symbiotic system, including 
 a 1.2$-$$1.4\,M_\odot$ WD and a 0.68$-$$0.8\,M_\odot$ RG star with an orbital period of $\sim$454\,d  (e.g. Brandi et al. 2009).
However, it is still unclear whether the WD in T CrB and  RS Oph is a CO WD or an ONe WD. 
Miko\l{}ajewska \& Shara (2017) recently argued that the WD in RS Oph may be a CO WD by analyzing its spectra, 
making it a  likely progenitor candidate of type Ia SNe.

In order to confirm the existence of AIC process, it is  important to understand electromagnetic signatures 
of AIC events and  to identify them in transient surveys. Piro \& Thompson (2014)  suggested that a particularly 
strong signature of an AIC event would occur for an ONe WD that  accretes material from a RG star. 
In such cases, the $\sim$$10^{50}$\,erg explosion from the AIC collides 
with  the surface of the RG companion, creating an X-ray flash lasting $\sim$1\,hr followed by an optical signature of AIC. 
Piro \& Thompson (2014)  argued that the strongest signal of the optical and X-ray emission will  come directly from the shocked region  though 
the strength of the signal would be strongly dependent on the viewing angle. 
Moriya (2016) suggested that the AIC radio transients originated from SD systems may be detected
in future radio transient surveys such as the Square Kilometer Array transient survey
and the Very Large Array Sky Survey. 
Future observational surveys may finally detect such events like AIC events with low luminosities (e.g. Piro \& Thompson 2014).

\section{Summary}

AIC is a theoretically  predicted final fate of WDs in stellar evolution, but there has never been a direct detection for such an event. 
In this work, I investigated the different SD channels of AIC events in a systematic way. 
I gave the initial parameter space of different SD channels based on detailed binary evolution computations, 
including the CO WD+He star channel and the ONe WD+MS/RG/He star channels.
According to a BPS study, I estimate that
the rates of AIC events in our galaxy are $\nu_{\rm AIC}\sim0.3-0.9\times10^{-3}$\,yr$^{-1}$, and that 
their delay times are longer than 30\,Myr. I found that  the ONe WD+He star channel is the main way to form AIC events, and that 
the CO WD+He star channel cannot be ignored when studying AIC events.
In order to provide constraints on the SD model of AIC events, 
large samples of observed WD binaries are expected.
Additionally, more numerical simulations related to the collapse of WDs  
are required to constrain the properties of the resulting AIC events,
and AIC events are expected to be identified and observed in nature by future surveys.

\section*{Acknowledgments}
BW acknowledges the anonymous referee for the valuable comments that helped him
improve this paper.   BW also thanks Zhanwen Han,  Philipp Podsiadlowski,  Zhengwei Liu and Takashi Moriya for their helpful discussions.
This study is supported by the National Natural Science Foundation of China (Nos 11873085, 11673059, 11390374  and 11521303),
the Chinese Academy of Sciences (Nos QYZDB-SSW-SYS001 and KJZD-EW-M06-01),
and the Yunnan Province (Nos 2017HC018 and 2018FB005).

\label{lastpage}
\end{document}